\documentclass[twocolumn,superscriptaddress,showpacs,amsmath,amssymb,pra,longbibliography]{revtex4-1}

\usepackage{graphicx} 
\usepackage{dcolumn}  
\usepackage{bm}       
\usepackage[usenames,dvipsnames]{color}
\definecolor{URLCOL}{rgb}{0,0.52,0.83} 
\definecolor{LINKCOL}{rgb}{0.05,0.5,0} 
\definecolor{orange}{rgb}{0.6,0.3,0} 
\definecolor{CITECOL}{rgb}{0.25,0,0.48} 
\usepackage{epstopdf}

\usepackage[bookmarks,bookmarksopen,bookmarksnumbered,colorlinks,linkcolor=LINKCOL,linktocpage,citecolor=CITECOL,urlcolor=URLCOL,pdfpagemode=UseOutlines]{hyperref}

\usepackage{booktabs}
\usepackage{comment}
\usepackage{natbib}

\definecolor{TITLECOL}{rgb}{0.1,0.2,0.7} 
\definecolor{SECOL}{rgb}{0.1,0.2,0.7} 
\definecolor{CONTENTSCOL}{rgb}{0.1,0.2,0.7} 
\definecolor{SSECOL}{rgb}{0.25,0,0.48} 
\definecolor{SSSECOL}{rgb}{0.2,0.08,0.53} 
\definecolor{FINCOL}{rgb}{0.01,0.3,0.07} 

\def\coloredtitle#1{\title{\textcolor{TITLECOL}{#1}}} 
\def\coloredauthor#1{\author{\textcolor{CITECOL}{#1}}} 

\definecolor{URLCOL}{rgb}{0,0.17,0.43} 
\definecolor{LINKCOL}{rgb}{0.05,0.4,0} 
\definecolor{CITECOL}{rgb}{0.35,0,0.48} 

\newcommand\BGid{\par\vskip 3ex
	\hspace{1.55cm}\raggedright{\sf\bfseries BurkeID:}\slshape\space}
\def\sss{\scriptscriptstyle\rm}

\def\bea{\begin{eqnarray}}
\def\eea{\end{eqnarray}}
\def\ben{\begin{equation}}
\def\een{\end{equation}}
\def\benu{\begin{enumerate}}
\def\enu{\end{enumerate}}

\def\bei{\begin{itemize}}
\def\eei{\end{itemize}}
\def\beit{\begin{itemize}}
\def\eit{\end{itemize}}
\def\benu{\begin{enumerate}}
\def\enu{\end{enumerate}}

\def\br{{\bf r}}

\def\half{\frac{1}{2}}

\def\x{_{\sss X}}

\def\xc{_{\sss XC}}

\def\GGA{^{\rm GGA}}

\def\unif{^{\rm unif}}

\def\sec#1{\section{\textcolor{SECOL}{#1}}}
\def\ssec#1{\subsection{\textcolor{SSECOL}{#1}}}

\def\n{\rho}

\def\dens{density}
\def\nucatt{{ NucAtt}}
\def\Hartree{{CoulRep}}
\def\orbKE{{OrbKE}}
\def\Xopt{{X^{\rm opt}}}
\def\elecrep{{ElecRep}}
\def\gradn{|\nabla\n|}

\begin{document}


\newcommand*\doctitle{DFT: A Theory Full of Holes?}

\def\ssec#1{{\em{#1}:}}
\def\sssec#1{{\em{#1}:}}

\coloredtitle{\doctitle}

\coloredauthor{Aurora Pribram-Jones}
\affiliation{Department of Chemistry,
University of California, 1102 Natural Sciences 2,
Irvine, CA 92697-2025, USA}

\coloredauthor{David A. Gross}
\affiliation{Department of Chemistry,
University of California, 1102 Natural Sciences 2,
Irvine, CA 92697-2025, USA}

\coloredauthor{Kieron Burke}
\affiliation{Department of Chemistry,
University of California, 1102 Natural Sciences 2,
Irvine, CA 92697-2025, USA}

\date{\today}

\begin{abstract}
This article is a rough, quirky overview of both the history and
present state of the art of density functional theory.  The field is so huge that
no attempt to be comprehensive is made.   We focus
on the underlying exact theory, the origin of approximations, and the 
tension between
empirical and non-empirical approaches.  
Many ideas are illustrated on the exchange energy and hole.
Features unique to this article include
how approximations can be systematically derived in a non-empirical
fashion and a survey of warm dense matter.  
\BGid{BG00101}
\vskip 0.5cm 
{\centering
\includegraphics[width=0.4\textwidth]{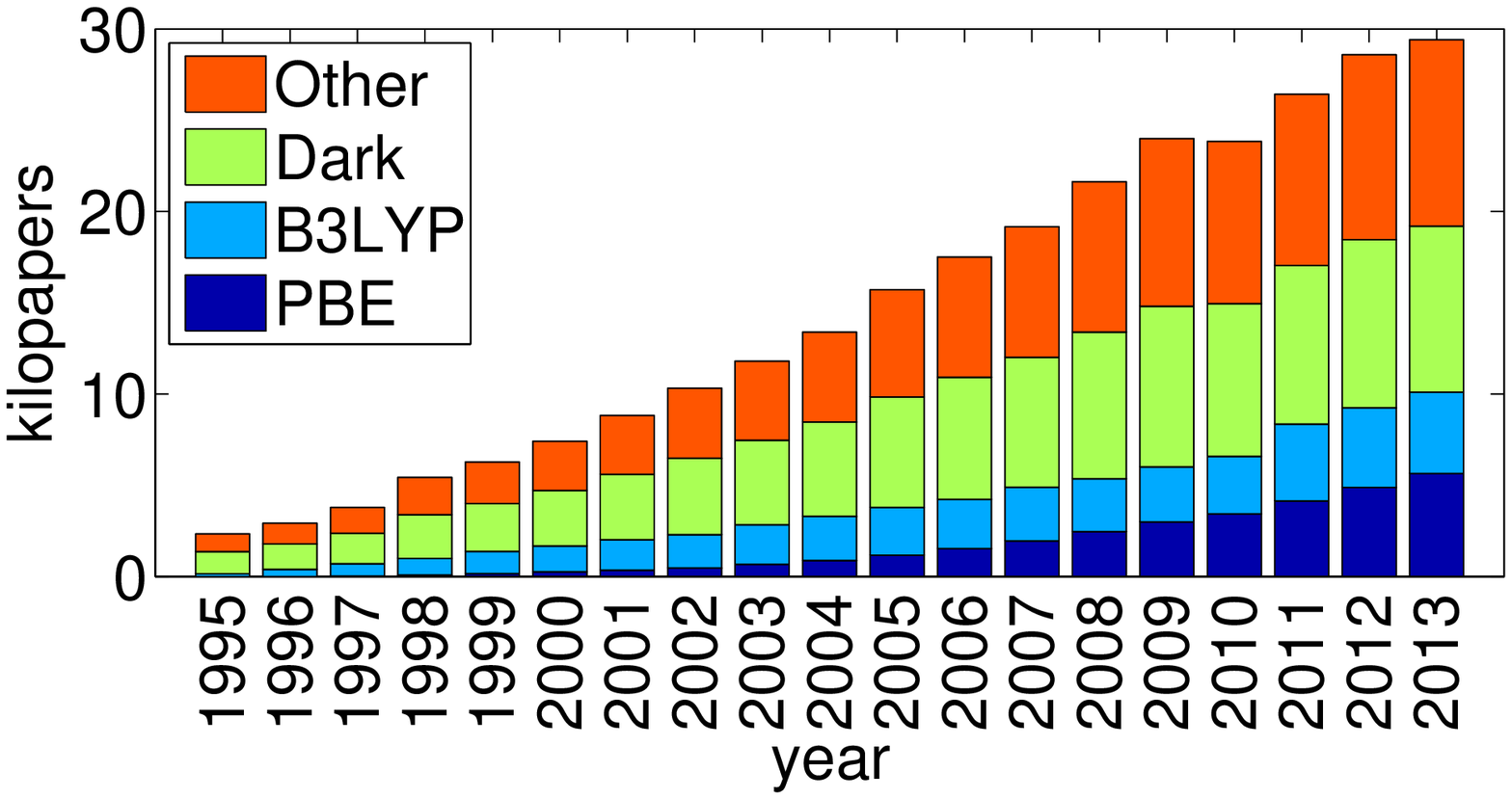}
\vskip 0.01cm
\begin{center}
The number of DFT citations has
exploded (as have {\em ab initio} methods).
PBE is the number of citations of Ref. \cite{PBE96}, 
and B3LYP of Ref. \cite{B93}.  {\em Dark} indicates papers
using either of these approximations without citing the original papers,
while {\em other} is all other DFT papers.  All numbers are estimates.
Contrast with Fig. 1 of Ref. \cite{B12}, which missed almost 2/3 of these.
\end{center}
}
\end{abstract}


\maketitle

\sec{What is this article about?}
\label{sec:intro}
The popularity of density functional theory (DFT) as an electronic structure method is unparalleled,
with applications that stretch from biology\cite{WSML13} to exoplanets\cite{KDLM12}.  
However, its quirks of logic and diverse modes of practical application have led
to disagreements on many fronts and from many parties.  
Developers of DFT are guided by many different principles, while
applied practitioners (a.k.a. users) are suspicious of DFT for
reasons both practical (\emph{how can I pick a functional with so many choices?}\cite{RCFB09}) and
cultural (\emph{with so many choices, why would I call this first-principles?}).  

A modern DFT calculation\cite{BW12} begins with the purchase of a computer, which might be as small as 
a laptop, and a quantum chemical code. Next, a basis set is chosen, which assigns predetermined functions to describe the electrons
on each atom of the molecule being studied.  Finally, a DFT approximation to something called
the exchange-correlation energy (XC) is chosen, and the code starts running.  For each guess
of the nuclear positions, the code calculates an approximate energy\cite{BW12}.  A geometry optimization
should find the minimum energy configuration.  With variations on this theme\cite{PY89,DG90}, one can read out
all molecular geometries, dissociation energies, reaction barriers, vibrational frequencies,
etc.  A modern desktop may do a calculation for a 100-atom system within a day.
A careful user will repeat the most important parts of the calculation with bigger basis sets,
to check that answers don't change significantly.

\sec{Where does DFT come from?}
\label{sec:origin}
Although DFT's popularity has skyrocketed since 
applications to chemistry became useful and routine, 
its roots stretch back much further\cite{B12,J12,Z14}.

\ssec{Ye olde DFT} 
Developed without reference to the
Schr{\"o}dinger equation\cite{S26}, 
Thomas-Fermi (TF) theory\cite{T27,F27,F28}
was the first DFT. It is pure DFT, relying only on
the electronic density, $\n(\br)$, as input.  
The kinetic energy was approximated as that of a uniform
electron gas, while the repulsion of the electrons
was modeled with the classical electrostatic Coulomb repulsion, 
again depending only on the electronic density as an input.  

\ssec{Mixing in orbitals}
John Slater was a master of electronic structure whose work foreshadowed the development of DFT.  In particular, his $X_\alpha$ method\cite{S51}
approximates the interactions of electrons in ground-state systems and improved upon Hartree-Fock (HF) \cite{F30,H35}, one of the simplest ways to capture the Pauli exclusion principle.  One of Slater's great insights was the importance of \emph{holes}, a way of describing the depressed probability of finding electrons close to one another.  Ahead of his time, Slater's $X_\alpha$ included
focus on the hole, satisfied exact conditions like sum rules, and considered of the degree of localization present in the system of interest.  

\ssec{A great logical leap}
Although Slater's methods provided an improvement upon HF, it was not until
1964 that Hohenberg and Kohn formulated their famous theorems\cite{HK64}, which now serve
as the foundation of DFT:

\noindent (i) the ground-state properties of an
electronic system are completely determined by $\n(\br)$, and 

\noindent (ii) there is a one-to-one correspondence between the external potential and the density.  

We write this by splitting the energy into two pieces:
\ben
E_{\rm elec}[\dens] = F[\dens] + \nucatt,
\een
where
$E_{\rm elec}$ is the total energy of the electrons, 
$F$ is the sum of their exact quantum kinetic and electron-electron
repulsion energies, and $\nucatt$ is 
their attraction to the nuclei in the molecule
being calculated.
Square brackets [ ] denote some (very complex) dependence on the 
one-electron density, $\n(\br)$, which gives the relative
probability of finding an electron in a small chunk of space around the
point $\br$.
$F$ is the same for all electronic systems, and so is called universal.  
For any given  molecule, your computer simply finds $\n(\br)$ that minimizes $E_{\rm elec}$
above.  
Compare this to the variational principle in regular quantum mechanics.
Instead of spending forever searching lots of wavefunctions
that depend on all $3N$ electronic coordinates, you
just search over one-electron densities, which have only 3 coordinates (and spin).

The pesky thing about the Hohenberg-Kohn theorems, however, is that they tell
us such things exist without telling us how to find them.
This means that to actually use DFT, we must approximate $F[\dens]$.
We recognize that the old TF theory did precisely this, with very crude
approximations for the two main contributions to $F$:
\ben
\label{eq:tf}
F[\dens] \sim \int d^3r\, \n^{5/3}(\br) + \Hartree,~~~~~~~(TF)
\een
where we've not bothered with constants, etc.  The first term is an approximation
to the kinetic energy as a simple integral over the density.  It is a local approximation, since the contribution at any point comes from only the density
at that point.  The other piece is the self-repulsion among electrons, which is
simply modeled as the classical electrostatic repulsion, often called their Hartree
energy or the direct Coulomb energy.
Such simple approximations are typically good to within about 10\% of the electronic
energy, but bonds are a tiny fraction of this, and so are not accurate in such
a crude theory\cite{T62}.

\ssec{A great calculational leap}
Kohn and Sham proposed rewriting the universal functional in order to approximate only a small piece of the energy. They mapped the interacting electronic system to a fake non-interacting system with the same $\n(\br)$.  
This requires changing the external potential, so these aloof, 
non-interacting electrons produce the same density as their interacting cousins.  The
universal functional can now be broken into new pieces.  Where in the interacting system,
we had kinetic energy and electron-electron interaction
terms, in the Kohn-Sham (KS) system, we write
the functional
\ben
\label{eq:ks}
F=\orbKE + \Hartree  + XC
\een
where $\orbKE$ is the kinetic energy of the fake KS electrons.
$XC$ contains all the rest, which includes both
kinetic and potential pieces. Although it is 
small compared to the total, `nature's glue' \cite{KP00} is critical
to getting chemistry and physics right. 
The $X$ part is (essentially) the Fock exchange from a HF calculation,
while $C$ is the correlation energy, i.e., that part that traditional
methods such as coupled cluster usually get very accurately\cite{BM07}.

\begin{figure}[htbp]
\includegraphics[width=\columnwidth]{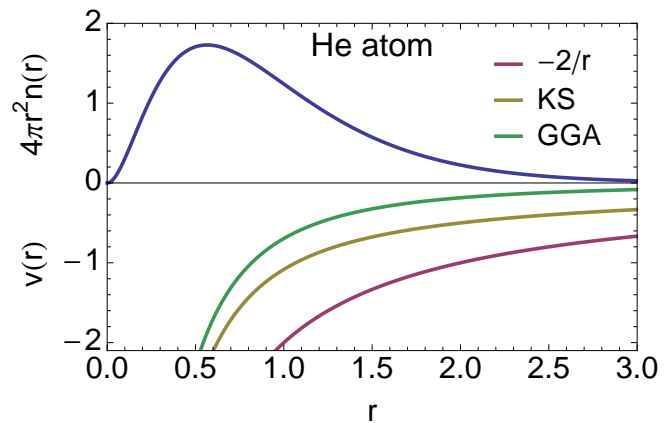}
\caption{
Radial densities and potentials for the He atom (energies in Hartree, distances
in Bohr).  The pink line is $-2/r$, the attraction of real electrons
to the nucleus.  The yellow is the {\em exact} KS potential. Two
fake electrons in the $1s$ orbital of this potential have the same
ground-state density as real He.  The green is the potential
of a typical approximation which, although inaccurate, yields a
highly accurate density.}
\label{fig:HedenpotLDA}
\end{figure}

When you minimize this new expression for the energy, you find
a set of orbital equations, the celebrated KS equations.
They are almost identical
to Hartree-Fock equations, and this showed that Slater's idea could be made
exact (if the exact functional were known).  
The genius of the KS scheme is that, because it calculates orbitals
and gives their kinetic energy, only $XC$, a small fraction of the total energy,
needs to be approximated as a density functional. 
The KS scheme usually produces excellent
self-consistent densities, even with simple approximations like
LDA, but approximate potentials
for this non-interacting KS system are typically
very different from the exact KS potential (Fig. \ref{fig:HedenpotLDA}).

\ssec{Popular approximations for XC}
Despite the overwhelming number of approximations available in the average DFT code, most calculations rely on a few of the most popular approximations.  The sequence of these approximations is
\bea
\label{eq:popxc}
XC&\sim& XC\unif(\n)~~~~~~~~~~~~~~~~~~~~~~~~~~~~(LDA)\nonumber\\
&\sim& XC\GGA(\n, \gradn)~~~~~~~~~~~~~~~~~~~~(GGA)\nonumber\\
&\sim& a(X-X\GGA)+ XC\GGA~~~~~~~~(hybrid)
\eea
The first was the third major step from the mid-60s and was invented in the KS paper\cite{KS65}.
It was the mainstay of solid-state calculations for a generation, and remains popular for
some specific applications even today.  It is (almost) never used in quantum chemistry, as
it typically overbinds by about 1eV/bond. The local density approximation (LDA)\cite{KS65} assumes that the XC energy depends on the density at each position only, and that dependence is the same as in a uniform electron gas.

Adding another level of complexity leads to the 
more accurate generalized gradient approximations (GGAs)\cite{P86,B88}, 
which use information about both the density and its
gradient at each point.  
Hybrid approximations mix a fraction ($a$) of 
exact exchange with a GGA\cite{B93}.
These maneuvers beyond the GGA usually increase
the accuracy of certain properties with an affordable
increase in computational cost\cite{PS01}.
(Meta-GGAs try to use a dependence on the KS kinetic energy density
to avoid calculating the Fock exchange of hybrids\cite{PK03,TPSS03},
which can be very expensive for solids.)  

\begin{figure}[htbp]
\includegraphics[width=\columnwidth]{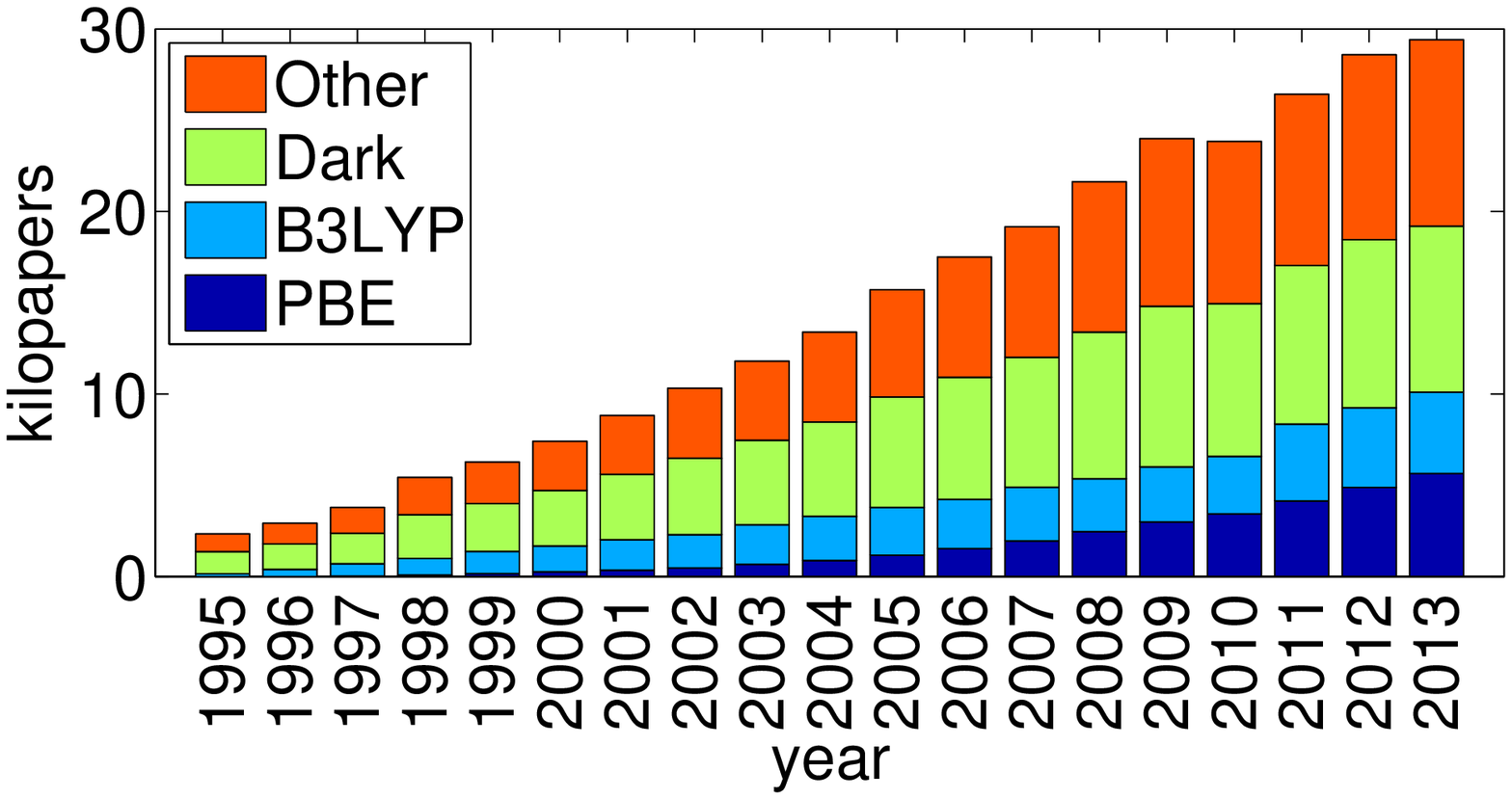}
\caption{The number of DFT citations has
exploded (as have {\em ab initio} methods).
PBE is the number of citations of Ref. \cite{PBE96}, 
and B3LYP of Ref. \cite{B93}.  {\em Dark} indicates papers
using either of these approximations without citing the original papers,
while {\em other} is all other DFT papers.  All numbers are estimates.
Contrast with Fig. 1 of Ref. \cite{B12}, which missed almost 2/3 of these.
}
\label{fig:kPapersPlot}
\end{figure}

Fig. \ref{fig:kPapersPlot} shows that the
two most popular functionals, PBE\cite{PBE96,PBE98} and B3LYP\cite{LYP88,B93}, 
comprise a large fraction of DFT citations each year (about 2/3), even though
they are now cited only about half the time they are used.
PBE is a GGA, while B3LYP is a hybrid\cite{B93}. 
As a method tied to Hartree-Fock, quantum chemists' old stomping grounds, and one
with typically higher accuracy than PBE, B3LYP is more often a chemist's choice. 
PBE's more systematic errors,  mathematical rationale, and
lack of costly exact exchange,  have made it most popular in solid-state
physics and materials science.  In reality, both are used in both
fields and many others as well. 

\ssec{Cultural wars}
\label{relwar}
The LDA was defined by Kohn and Sham in 1965; there is no controversy
about how it was designed.  
On the other hand, adding complexity to functional approximations
demands choices about how to take the next step. 
Empirical functional developers fit their approximations to
sets of highly accurate reference data on atoms and molecules. 
Non-empirical developers use exact mathematical conditions on the
functional and rely on reference systems like the uniform and slowly-varying electron gases.
The PBE GGA is the most popular non-empirical approximation, while the most popular empirical functional approximation is the B3LYP hybrid. Modern DFT conferences usually include debates about the morality of this kind of empiricism.

Both philosophies have been incredibly successful, 
as shown by their large followings among developers and users, 
but each of these successes is accompanied by failures. 
 No single approximation works well enough for every property 
of every material of interest.  Many users sit squarely
and pragmatically in the middle of the two factions, 
taking what is best from both of their accomplishments and insights.  
Often, empiricists and non-empiricists 
find themselves with similar end products, 
a good clue that something valuable has been created with the strengths of both.

\begin{figure}[htbp]
\includegraphics[width=\columnwidth]{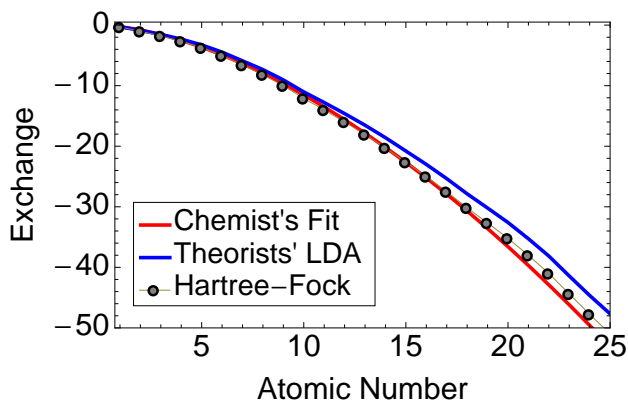}
\caption{Exchange energy (in Hartrees) of atoms from a HF calculation as a function
of $Z$, atomic number, and two LDA X 
calculations, one with the theoretical asymptote, the other fitted.}
\label{fig:XvsZ}
\end{figure}

To illustrate this idea, we give a brief allegory from an alternative
universe.  Since at least the 1960s, accurate HF energies of atoms
have been available due to the efforts of Charlotte Froese Fischer and
others\cite{F69,F77}.  A bright young chemistry student plots these X energies as a function of 
$Z$, the atomic number, and notices they behave roughly as $Z^{5/3}$,
as in Fig. \ref{fig:XvsZ}.
She's an organic chemistry student, and mostly only
cares about main-group elements, so she fits the curve by choosing a
constant to minimize the error on the first 18 elements, finding 
$E\x=-0.25 Z^{5/3}$.  Much later, she hears about KS DFT, and the need
to approximate the XC energy.  A little experimentation shows that if
\ben
\Xopt = C_0 \int d^3 r~\n^{4/3}(\br),
\een
this goes as $Z^{5/3}$ when $Z$ is large, and choosing $C_0=-0.80$ makes
it agree with her fit.

In our alternate timeline, a decade later, Paul Dirac, a very famous physicist,
proves\cite{D30} that for a uniform gas, $C_0=A\x=-(3/4)(3/\pi)^{1/3}=-0.738$.  Worse still,
Julian Schwinger proves\cite{S81} that inserting the TF density into Dirac's expression
becomes exact as $Z\to\infty$, so that $E\x\to -0.2201 Z^{5/3}$.   
Thus theirs is the `correct' LDA for X, and our brave young student should
bow her head in shame.  

Or should she?  If we evaluate the mean absolute errors in exchange
for the first 20 atoms, her functional is significantly better than the
`correct' one\cite{HC01}.  If lives depend on the accuracy for
those 20 atoms, which would you choose\cite{note1}?

This simple fable contains the seeds of our actual cultural wars in DFT
derivations:

\noindent
(i) An intuitive, inspired functional need not wait for an official derivation.
One parameter might be extracted by fitting, and later derived.

\noindent
(ii) A fitted functional will usually be more accurate than the derived
version for the cases where it was fitted.  The magnitude of the
errors will be smaller, but less systematic.

\noindent
(iii) The fitted functional will miss universal properties of
a derived functional.  We see in Sec. \ref{sec:sysapprox} that the correct LDA for exchange
is a universal limit of {\em all} systems, not just atoms.

\noindent
(iv) If you want to add the next correction to LDA, starting with the
wrong constant will make life very difficult (see Sec. \ref{sec:sysapprox}).

\begin{figure}[htbp]
\includegraphics[width=\columnwidth]{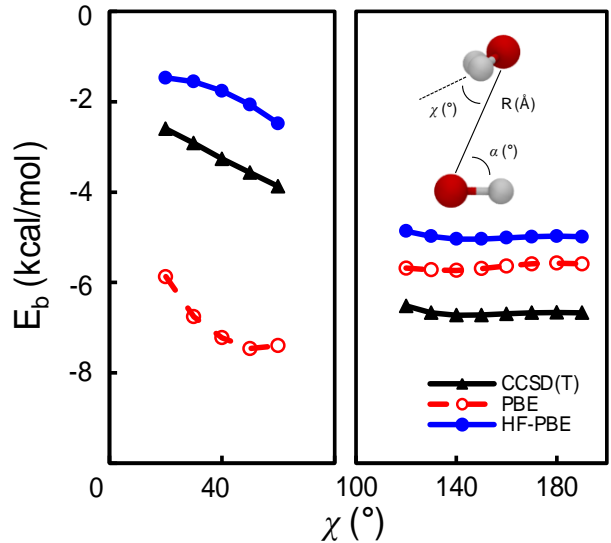}
\caption{When a DFT calculation is {\em abnormally} sensitive
to the potential, the density can go bad.
Usually, DFT approximate densities are better than HF\cite{BGM13}, as in Fig 1.
Here, self-consistent PBE results 
for
${\text OH-H}_2{\text O}$ interactions yield the wrong geometry,
but PBE on HF-densities fixes this\cite{KSB14}.}
\label{fig:HFPBESketch}
\end{figure}

\sec{What's at the forefront?}
\label{sec:fore}
\ssec{Accurate Gaps}
Calculating accurate energy gaps and self-interaction
errors are notorious difficulties within DFT\cite{FNM03}.  
Self-interaction error (SIE) stems from spurious interaction of an
electron with itself in the Coulomb repulsion term.  
Orbital-dependent methods often cure most of the SIE problem, 
but they can be expensive to run.  
The `gap problem' in DFT often stems from treating
the KS HOMO-LUMO gap as the fundamental gap, 
but the difference in the HOMO and LUMO of the
KS system is not the same as the
difference between the ionization potential 
and the electron affinity\cite{FNM03}.  
Ad hoc methods are often used to correct DFT gaps, 
but these methods require expensive additional calculations, 
empirical knowledge of your system, or empirical tuning.  
However, it has recently been shown that some classes of
self-interaction error are really just errors due to 
poor potentials leading to poorer densities
\cite{KSB13,KSB14}.  
Such errors are removed by using more accurate densities (Fig. \ref{fig:HFPBESketch}). 

\ssec{Range-separated hybrids}
Range-separated hybrids\cite{HSE06} improve fundamental gaps calculated via the DFT HOMO-LUMO gap\cite{KSRB12}. Screened range-separated hybrids can even achieve gap renormalization present when moving between gas-phase molecules and molecular crystals\cite{RSJB13}.  The basic range-separated hybrid scheme separates the troublesome Coulomb interaction into long-range and short-range pieces. The screened version enforces exact conditions to determine where this separation occurs and incorporates the dielectric constant as an adaptive parameter. This technique takes into account increased screening as molecules form solids, resulting in reduced gaps critical for calculations geared toward applications in molecular electronics.

\ssec{Weak Interactions}
\label{sec:weak}
Another of DFT's classic failings is its poor treatment of weak interactions\cite{JG89,GDP96}.  
Induced dipoles and the resulting dispersion interactions are not captured by the most popular approximations
of Eq. \ref{eq:popxc}. This prevents accurate modeling
of the vast majority of biological systems, as well as a wide range
of other phenomena, such as surface adsorption and molecular crystal packing. GGAs and
hybrids are unable to model the long-range correlations occurring between fluctuations induced 
in the density.  The non-empirical approach based on the work of
Langreth and Lundqvist\cite{ALL96,DRSL04,SAGG02,LKBA11} and the
empirical DFT-D of Grimme\cite{G06,JSCH06} have dominated the advances in this area, 
along with the more recent, less empirical approach of Tkatchenko and Scheffler\cite{TS09,ZTPA11}.

\sec{Reducing cost: Is less more?}
\label{sec:cost}
No matter how much progress is made in improving algorithms to
reduce the computational cost of DFT calculations, there will
always be larger systems of interest, and even the fastest calculations become prohibitively expensive.  
The most glaring example is molecular dynamics (MD) simulation in biochemistry.  With classical
force fields, these can be run for nano- to milli-seconds, with a million
atoms, with relative ease.   But when bonds break, a quantum treatment
is needed, and the first versions of these were recognized in last year's
Nobel prize in chemistry\cite{WL76,Lb01,K06}.  These days, many people run Car-Parrinello
MD\cite{CP85,IMT05}, with DFT calculations inside their MD, but this reduces tractable
system sizes to a few hundred atoms.

Because of this, there remains a great deal of interest in finding clever
ways to keep as much accuracy as needed while simplifying computational steps.  One method for doing so involves circumventing the
orbital-dependent KS step of traditional DFT calculations.  
Alternatively, one can save time by only doing those costly
steps (or even more expensive procedures) on a system's
most important pieces, while leaving the rest to be calculated
using a less intensive method. The key to both approaches
is to achieve efficiency without sacrificing precious accuracy.

\ssec{Removing the orbitals}
Orbital-free methods\cite{DG90,WC00,KJTH09,KJTHb09,KT12,SRHM12} 
like TF
reduce computational costs, but are
often not accurate enough
to compete with KS DFT calculations.  
Current methods search for a
similar solution, by working on non-interacting kinetic energy
functionals that allow continued use of existing XC functionals\cite{KH03}.
(An intriguing alternative is to use the potential as the basic
variable
\cite{CLEB11,CGB13} -- see Secs. \ref{sec:sysapprox} and \ref{sec:wdm}.)

\ssec{Embedding}
Partitioning and embedding are similar procedures, in
which calculations on isolated pieces of a molecule are used
to gain understanding of the molecule as a whole\cite{LT07}.  One might want
to separate out molecular regions to
look more closely at pieces of high interest or to find
a better way to approximate the overall energy with density functionals.  
Parsing a molecule into chunks can also allow for entirely new
computational approaches not possible when dealing with the molecule as a whole.

Partition DFT\cite{EBCW10} is an exact embedding method based on density
partitioning\cite{CW06,CW07}.  Because it uses ensemble
density functionals\cite{GOK88,PYTB14}, it can handle non-integer
electron numbers and spins\cite{TNW12}. Energy of the fragments is minimized using effective potentials consisting of a fragment's potential and a global partition potential that maintains the correct total density.  This breakdown into fragment and partition energies allows approximations good for localized systems to be used alongside those better for the extended effects associated with the partition potential.

While partition DFT uses DFT methods to break up the system,
projector-based wavefunction-theory-in-DFT embedding techniques combine wavefunction and DFT methods\cite{MSGM12,BGMM13}.  This multiscale approach leverages the increased accuracy of some wavefunction methods for some bonds, where high accuracy is vital, without extending this computational cost to the entire system.  Current progress in this field has been toward the reduction of the errors introduced by the mismatch of methods between subsystems. This type of embedding has been recently applied to heterolytic bond cleavage and conjugated systems\cite{GBMM14}. Density matrix embedding theory on lattices\cite{KC12} and its extension to full quantum mechanical chemical systems\cite{KC13} use ideas from the density matrix renormalization group (DMRG)\cite{W92,W93}, a blazingly fast way to exactly solve low-dimensional quantum mechanics problems. This shifts the interactions between fragments to a quantum bath instead of dealing with them through a partition potential. 

\sec{What is the underlying theory behind DFT approximations?}
\label{sec:under}
Given the Pandora's box of approximate functionals, many 
found by fitting energies of systems, most users imagine DFT as an
empirical hodgepodge.  Ultimately, if we end up with a different functional
for every system, we will have entirely defeated the idea of first-principles
calculations. However, prior to the mid-90s, many decades of theory 
were developed to better understand the local approximation and
how to improve on it\cite{JG89}.  Here we summarize the most
relevant points.

The joint probability of finding one electron in 
a little chunk of space around point $A$ and another in some other chunk of space around point $B$ is
called the pair probability density.  The exact quantum repulsion among electrons is then
\ben
\elecrep =\half \int dA \int dB\, \frac{P(A,B)}{|\br_A - \br_B|}.
\label{elecrep}
\een
But we can also write
\ben
P(A,B)=\n(A)\, \n_\text{cond}(A,B).
\een
where $\n(A)$ is the density at $\br_A$ and $\n_\text{cond}(A,B)$ is the probability of finding the second
electron at B, {\em given} that there's one at A.  (If you ignore the electron at A, this is just $\n(B)$,
and Eq. \ref{elecrep} gives the Coulomb repulsion in Eqs. \ref{eq:tf} and \ref{eq:ks}).  We write this conditional probability as
\ben
\n_\text{cond}(A,B)=\n(B)+\n\xc(A,B).
\een
where $\n\xc(A,B)$ is called the hole around A.  It is mostly negative and represents a missing electron
(it integrates to -1), since the conditional probability integrates to $N-1$.  
With a little math trick (called the adiabatic connection\cite{LP75,GL76}), we can 
fold the kinetic correlation
into the hole so that
\ben
\label{XChole}
XC =\half \int dA \int dB\, \frac{\n(A)\, \n\xc(A,B)}{|\br_A - \br_B|}.
\een

\begin{figure}[htbp]
\includegraphics[width=\columnwidth]{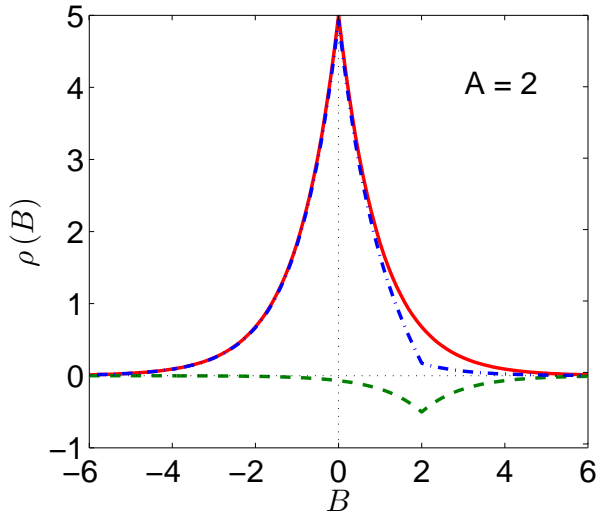}
\caption{Cartoon of a one-dimensional 10-electron
density (solid red), the conditional density (dot-dashed blue) given
an electron at $A=2$, and its hole density (dashed green).}
\label{fig:HoleCartoon}
\end{figure}
Because the XC hole tends
to follow an electron around, i.e., be centered on $A$ as in Fig. \ref{fig:HoleCartoon}, its shape is 
roughly a simple function of $\n(A)$.  If one approximates the hole by that
of a uniform gas of density $\n(A)$, Eq. \ref{XChole} above yields the LDA for the XC energy.
So the LDA approximation for XC can be
thought of as approximating the hole by that of a uniform gas\cite{JG89,EBP96}.  

\begin{figure}[htbp]
\includegraphics[width=\columnwidth]{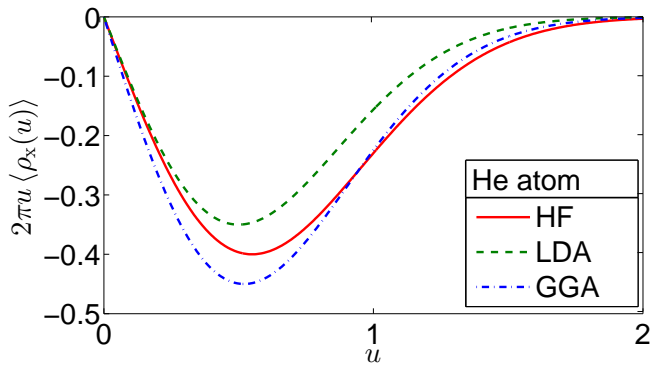}
\caption{Representation of system-averaged  radial exchange holes
for the helium atom\cite{EP98},
weighted by the Coulomb repulsion, so that
the area equals the X energy.
The LDA hole (dashed green) is not deep enough, reflecting
the LDA underestimate of the magnitude of the X energy.
The GGA hole (dot-dashed blue) is substantially better, but a little
too deep.}
\label{fig:XholeGGA}
\end{figure}
But while the XC is roughly approximated by LDA, 
the energy density at each point in a system is not,
especially in systems of low symmetry. But from Eq. \ref{XChole}, 
the energy depends only on the {\em average} of the XC hole
over the system, and Fig. \ref{fig:XholeGGA} shows such a system-averaged hole for the He atom.
(Integrate over $A$ and the angular parts of $B$ in Eq. (\ref{XChole}).)
The LDA hole
is not deep enough, and neither is the LDA energy.  This is the effect that leads to LDA
overbinding of molecules.

\ssec{GGA Made Briefer}
The underlying idea behind the Perdew series of GGAs was to improve on the LDA hole\cite{B97}.  Adding
gradient corrections to the hole violates certain sum rules (negativity of the X hole and
integration to -1, and integration to 0 for the correlation hole), so the real-space cutoff
procedure was designed to restore these conditions.  This is an effective resummation of
the gradient expansion, producing the numerical GGA.  The popular functional PBE was derived
from imposing exact conditions on a simple form\cite{PBE96,PBE98}, but should be believed because it mimics
the numerical GGA.  In Fig. \ref{fig:XholeGGA}, we show how the GGA hole roughly improves on LDA, reducing typical
energy errors by a factor of three.

GGAs don't only show how important good hole models can be.  They also
demonstrate that good approximations can satisfy different exact conditions,
so picking which to satisfy is non-trivial.  For instance, B88\cite{B88}, PW91\cite{PCVJ92,PCVJ93},
and PBE\cite{PBE96,PBE98} give similar values for exchange energy when densities
do not get too small or vary too quickly.  However, once they do, each behaves
very differently.  Each approximation was sculpted to satisfy different
exact conditions in this limit.  Becke decided a good energy density
for exponential electronic densities was important. Perdew first
thought that a particular scaling behavior was important\cite{BPW97}, then
that satisfying a certain bound was better\cite{PBE96}.  Without a
systematic way to improve our approximations, these difficult choices guide our progress.  
But starting from a model for the XC hole is an excellent idea, as such a model
can be checked against the exact XC hole\cite{CF12}.

\ssec{XDM}
A recent, parameter-free approach to capturing dispersion 
is the exchange-hole dipole moment (XDM) method\cite{JB06,BJ07,BJb07,OCPH14},
where perturbation theory yields
a multipole-multipole interaction, and quantum
effects are included through the dipole moment of
the electron with its exchange hole.  
Using these in concert with atomic polarizabilities
and dipole moments generates atomic pair dispersion
coefficients that are within 4\% of reference $C_6$ values\cite{BJ06}.
Such a model has the advantage over the more popular methods of Sec. \ref{sec:fore} because
its assertions about the hole can be checked.

\ssec{RPA and other methods}
Originally put forth in the 1950s as a method for the uniform electron gas, the random phase approximation (RPA) can be viewed as a simplified wavefunction method or a nonlocal density functional approach that uses both occupied and unoccupied KS states to approximate the correlation energy.   RPA correlation performs extremely well for noncovalent, weak interactions between molecules and yields the correct dissociation limit of $H_2$\cite{FNGB05}, two of the major failures of traditional DFT approximations\cite{CMY08}. 

Though computational expense once hindered its wide use, 
resolution-of-identity implementations\cite{F08,EYF10} have improved 
its efficiency, making RPA accessible to researchers interested in large molecular systems.  RPA gives good dissociation energy for catalysts involving the breaking of transition-metal-ligand and carbon-carbon bonds in a system of over 100 atoms\cite{EF11}.  Though RPA handles medium- and long-range interactions very well, its trouble with short-range correlations invites development of methods that go `beyond RPA.'  RPA used in quantum chemistry usually describes only the particle-hole channel of the correlation, but another recent approach to RPA is the particle-particle RPA (pp-RPA)\cite{AYY14}. pp-RPA is missing some correlation, which causes errors in total energies of atoms and small molecules. This nearly cancels out in reaction energy calculations and yields fairly accurate binding energies\cite{PSAY13}.

RPA and variations on it will likely lead to
methods that work for both molecules and solids, and their computational
cost will be driven down by algorithmic development. However, RPA is likely
to remain substantially more expensive than a GGA calculation for the
indefinite future.  While it may rise to fill an important
niche in quantum chemistry, producing comparably accurate energetics to modern functionals without
any empiricism, such methods will not {\em replace} DFT as the first
run for many calculations.
Moreover, as with almost all `better' methods than DFT, there appears
to be no way to build in the good performance of older DFT approximations. 


\sec{Is there a systematic approach to functional approximation?}
\label{sec:sysapprox}
A huge intellectual gap in DFT development has been in the theory
behind the approximations. This, as detailed above, has allowed the
rise of empirical energy fitting.  Even the most appealing non-empirical
development seems to rely on picking and choosing which exact conditions
the approximation should satisfy.   Lately, even Perdew has resorted to
one or two parameters in the style of Becke\cite{PKZB99,SHXB13}, in order to construct
a meta-GGA.  Furthermore, up until the mid 1990s, many
good approximations were developed as approximations to the XC hole,
which could then be tested and checked for simple systems.

However, in fact, there {\em is} a rigorous way to develop density
functional approximations. Its mathematical foundations were laid down 40 years
ago by Lieb and Simon\cite{LS73,LS77,L81}.
They showed the fractional error in the energy in any TF calculation
vanishes as $Z\to\infty$, keeping $N=Z$.
Their original proof is for atoms, but applies to any molecule or
solid, once the nuclear positions are scaled by $Z^{1/3}$ also.
Their innocuous statement is in fact very profound.  
This very complicated many-body quantum problem, 
in the limit of {\em large} numbers of electrons, has an almost
trivial (approximate) solution.  
And although
the world finds TF theory too inaccurate to be useful,
and performs KS calculations instead, 
the equivalent statement (not proven with rigor) is
that the fractional error in the LDA for XC vanishes as $Z \to \infty$.
XC, like politics, is entirely local in this limit.

These statements explain many of the phenomena we see in modern DFT:

\noindent
(i) LDA is {\em not} just an approximation that applies for uniform
or slowly varying systems, but is instead a universal limit of {\em all}
electronic systems.  

\noindent
(ii) LDA is the leading term in an asymptotic expansion in powers of
$\hbar$, i.e., semiclassical.  
Such expansions are notoriously difficult to deal with mathematically.

\noindent
(iii) The way in which LDA yields an ever smaller error as $Z$ grows
is very subtle.  The leading corrections are of several origins.
Often the dominant error is a lack of spatial quantum oscillations in the
XC hole.  However, as $Z$ grows, these oscillations get faster,
and so their net effect on the XC energy becomes smaller.  Thus,
even as $Z$ grows, LDA
should not yield accurate energy densities everywhere in a system
(and its potential is even worse, as in Fig. \ref{fig:HedenpotLDA}), but the integrated XC energy will become ever
more accurate.

\noindent
(iv) The basic idea of the GGA as the
leading correction to LDA makes sense.   The leading corrections
to the LDA hole should exist as very sophisticated functionals
of the potential, but whose energetic effects can be captured by
simple approximations using the density gradient.  This yields improved
net energetics, but energy densities might look even worse, especially
in regions of high gradients, such as atomic cores.

\begin{figure}[htbp]
\includegraphics[width=\columnwidth]{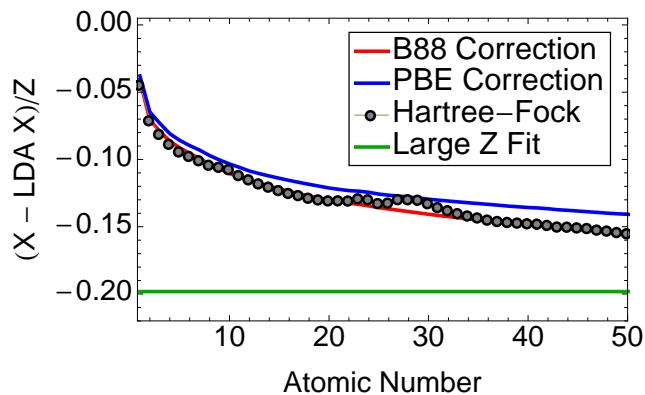}
\caption{The non-local exchange energy (exchange minus LDA X) per
electron of atoms with atomic number $Z$ (compare with Fig. \ref{fig:XvsZ}).
The PBE functional tends to the theoretical limit ($Z\to\infty$)
 (horizontal green
line), but B88 is more accurate for $Z < 50$ because of fitting\cite{EB09}.}
\label{fig:LDAXCorr}
\end{figure}

Next, we continue the allegory from Sec. \ref{relwar}.  To do so, we subtract
the LDA X energy from our accurate ones, so we can see the next correction,
and plot this, per electron,
in Fig. \ref{fig:LDAXCorr}.
Now, a bright young chemist has heard
about the GGA, cooks up an intuitive correction to LDA, and fits one parameter to the noble
gas values.  Later, some physicists derive a different GGA, which
happens to also give the correct value.   Later still, a derivation
of the correction for large $Z$ is given, which can be used to
determine the parameter (and turns out to match the empirical value
within 10\%).   The only difference from the original allegory is that
this is all true.  The chemist was Axel Becke; his fitted functional
is B88\cite{B88}. The derived functional is PBE\cite{PBE96}, and the derivation of the
parameter in B88 is given in Ref. \cite{EB09}.

This true story 
both validates Becke's original procedure and the
semiclassical approach to density functional approximation.
Note that even the correction is evaluated on the TF density
to find the limiting behavior.
The PBE exchange functional also yields the leading
the correction to the exchange energy of atoms.  By throwing this
away and restoring the (different) gradient expansion for slowly-varying
gases, PBEsol was created\cite{ES99}.

\ssec{Semiclassical approximations}
New approximations driven by semiclassical research can be divided into density approaches and potential approaches.  In the density camp, we find innovations like the approximations by Armiento and Mattsson\cite{AM05,MA09,MA10}, which incorporate surface conditions through their semiclassical approach. In the potential functional camp, we find highly accurate approximations to the density, which automatically generate approximations to non-interacting kinetic energies\cite{CLEB10,CLEB11,CGB13}.  
Since these approaches use potential functionals, they are
orbital-free and incredibly efficient, but only apply in one dimension (see also Sec. \ref{sec:wdm}).
Current research is focused on extension to three dimensions, semiclassical approximations in the presence of classical turning points, as well as semiclassical approximations to exchange and correlation energies.

\sec{Warm dense matter: A hot new area?}
\label{sec:wdm}
Though we do not live at icy absolute zero, most chemistry and physics happens at low enough temperatures that electrons are effectively in their ground state.  Most researchers pretend to be at zero temperature for their DFT work with impunity.  But some people, either those working at high enough temperatures and pressures or those interested in low-energy transitions, can't ignore thermal effects.  Those of us caught up in these warmer pursuits must tease out where temperature matters for our quantum mechanical work.

Mermin proved a finite-temperature version of the Hohenberg-Kohn theorem in 1965\cite{M65}, and the finite-temperature LDA was shown in the original KS paper\cite{KS65}.  However, many people continue to rely on the zero-temperature approximations, though they populate states at higher energy levels using finite-temperature weightings.  Better understanding and modeling of the finite-temperature XC hole could lead to improvement in some of the finer details of these calculations, like optical and electronic properties\cite{PPGB14}.

\ssec{WDM and MD}
One area that has seen great recent progress with DFT is the study of warm
dense matter (WDM)\cite{HEDP03,GDRT14}.  WDM is intermediate to solids and
plasmas, inhabiting a world where both quantum and classical
effects are important.  It is found deep within planetary interiors,
during shock physics experiments, and on the path to ignition
of inertial confinement fusion.  Lately, use of DFT MD has been
a boon to researchers working to simulate these complicated
materials\cite{MD06,HRD08,KRDM08,KD09,RMCH10}.  Most of these calculations are 
performed using KS orbitals with thermal occupations,
ignoring any temperature dependence of XC, 
in hopes that the kinetic and Coulomb energies
will capture most of the thermal effects.  
Agreement with experiment has been excellent, though there is
great interest in seeing if temperature-dependent XC approximations affects these results.

\begin{figure}[htbp]
\includegraphics[width=\columnwidth]{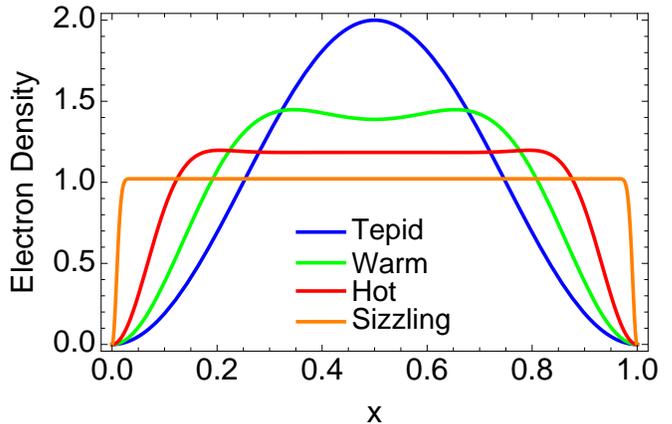}
\caption{The density of a single electron in a flat box
spreads toward the infinite walls as temperatures rise.}
\label{fig:boxden}
\end{figure}

\ssec{Exact conditions}
Exact conditions have been derived\cite{E10,PPFS11,DT11,PPGB14} for finite-temperature systems that seem very similar to their ground-state counterparts.  However, a major difference in thermal systems is that when one squeezes or compresses the length scale of the system, one sees an accompanying scaling of the temperature.  This is further reflected in the thermal adiabatic connection, which connects the non-interacting KS system to the interacting system through scaling of the electron-electron interaction.  At zero temperature, this allows us to write the XC energy in terms of the potential alone, as long as it is accompanied by appropriate squeezing or stretching of the system's length scale (see Sec. \ref{sec:under}).  With the temperature-coordinate scaling present in thermal ensembles, the thermal adiabatic connection requires not only length scaling, but also the correct temperature scaling.

\begin{figure}[htbp]
\includegraphics[width=\columnwidth]{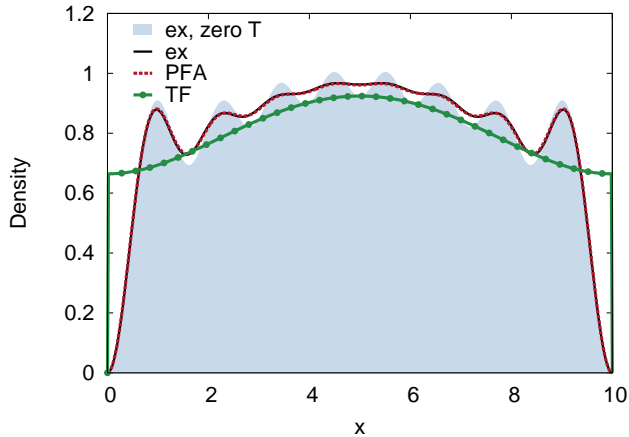}
\caption{Eight electrons in the potential $-2\sin^2{(\pi x/10)}$
in a 1d box.  At zero temperature (gray), 
the density exhibits sharp quantum oscillations, which wash out
as the 
temperature increases (black).  This effect is much weaker
near the edges.  TF is used in many warm simulations, 
but (green) misses all oscillations, vital for accurate chemical
effects.   The 
orbital-free, finite-temperature potential functional approximation of
Ref. \cite{CP14} is almost exact here (red).}
\label{fig:FTPFTexample}
\end{figure}

\ssec{OF Methods}
Orbital-free methods, discussed in Sec. \ref{sec:cost}, are of particular interest in the WDM community. Solving the KS equations with many thermally populated orbitals is repeated over and over in DFT MD, leading to prohibitive cost as temperatures rise.  The focus on free energies for thermal ensembles has led to two different approaches to orbital-free approximations.  One approach uses two separate forms for kinetic and entropic contributions\cite{DT11}. Following this path, one can either make approximations empirically\cite{KST12} or non-empirically\cite{KCST13}.  Another approach enforces a particular type of response in the uniform gas limit\cite{SD13}.  If one wishes to approximate the kentropy\cite{PPGB14} as a whole, one can use temperature-dependent potential functional theory to generate highly accurate approximations from approximate densities generated semiclassically or stochastically\cite{BNR13,CP14}.  Fig. \ref{fig:FTPFTexample} shows the accuracy of a semiclassical density approximation, which captures the quantum oscillations missed by Thomas-Fermi and still present as temperatures rise.

\sec{What can we guess about the future?}
\label{sec:guess}
The future of DFT remains remarkably bright.  As Fig. \ref{fig:kPapersPlot} shows,
the number of applications continues to grow
exponentially, with three times as much activity
than previously realized (Fig. 1 of \cite{B12}).   While empiricism has generated
far too many possible alternatives, the standard well-derived approximations
continue to dominate.

To avoid losing insight, it is important to further develop the
systematic path to approximations, which eschews all empiricism
and expands the functional in powers of $\hbar$, Planck's constant.
This will ultimately tell us what we can and cannot do with local-type
approximations.  There is huge room for development in this area,
and any progress could impact all those applications.

Meanwhile, new areas have been (e.g. weak interactions) or
are being developed (warm dense matter).  New methods, such as using
Bayesian statistics for error analysis\cite{MWVS14} or
machine learning for finding functionals\cite{SRHM12,SRHB13}, are coming
on line.  Such methods will not suffer the limitations of local
approximations, and should be applicable to strongly correlated
electronic systems, an arena where many of our present approximations
fail.  We have little doubt that DFT will continue to thrive for
decades to come.

\sec{Acknowledgments}
APJ thanks the U.S. Department of Energy (DE-FG02-97ER25308), and 
KB thanks the National Science Foundation (CHE-1112442).  
We are grateful to Cyrus Umrigar for data on the helium atom and to
Min-Cheol Kim  for Fig. \ref{fig:HFPBESketch} and Attila Cangi for
Fig. \ref{fig:FTPFTexample}.

\nocite{SK12}

\label{page:end}

\begin{thebibliography}{100}
\bibitem{note1}
In fact, sadly, the young chemist is unable to find a permanent position, and
  she ends up selling parametrized functionals for food on the streets. On the
  other hand, the physicists all celebrate their triumph over empiricism with a
  voyage on a brand new ship, which has been designed with materials whose
  properties have been calculated using DFT. Because the local approximation,
  as given above, underestimates the magnitude of the exchange energy, the
  brittle transition temperature is overestimated. When the new ship sails
  through icy waters, its hull is weakened and damaged by an iceberg, so all of
  them drown. (The interested reader may find more information on the
  ductile-to-brittle transition in Ref. \cite{SK12} and other works by
  Kaxiras.).
\bibitem{LT07}
H.~Lin and D.~Truhlar, ``{Q}{M}/{M}{M}: {w}hat have we learned, where are we,
  and where do we go from here?,'' {\em Theoretical Chemistry Accounts},
  vol.~117, no.~2, pp.~185--199, 2007.
\end{thebibliography}
\end{document}